%% file: PEAR.tex
\documentclass[sigconf]{acmart}
\AtBeginDocument{%
  \providecommand\BibTeX{{%
    \normalfont B\kern-0.5em{\scshape i\kern-0.25em b}\kern-0.8em\TeX}}}

\copyrightyear{2022}
\acmYear{2022}
\setcopyright{acmlicensed}
\acmConference[WWW '22 Companion]{Companion Proceedings of the Web Conference 2022}{April 25--29, 2022}{Virtual Event, Lyon, France}
\acmBooktitle{Companion Proceedings of the Web Conference 2022 (WWW '22 Companion), April 25--29, 2022, Virtual Event, Lyon, France}
\acmPrice{15.00}
\acmDOI{10.1145/3487553.3524208}
\acmISBN{978-1-4503-9130-6/22/04}

\usepackage{enumitem}
\usepackage{multirow}
\usepackage{amsmath}
\usepackage{balance}

\begin{document}
\fancyhead{}
\title{PEAR: Personalized Re-ranking with Contextualized Transformer for Recommendation}

\author{Yi Li$^{1\ast}$,$\:$ Jieming Zhu$^{2\ast}$,$\:$ Weiwen Liu$^2$,$\:$ Liangcai Su$^1$,$\:$ Guohao Cai$^2$,$\:$ Qi Zhang$^2$,$\:$ Ruiming Tang$^2$}
\author{Xi Xiao$^{1\star}$,$\:$ Xiuqiang He}

\affiliation{\vspace{1.5ex}
$^1$Shenzhen International Graduate School, Tsinghua University \country{China} \\
$^2$Huawei Noah's Ark Lab, China \\\vspace{0ex}
\texttt{\{yi-li20,} \texttt{sulc21\}@mails.tsinghua.edu.cn, xiaox@sz.tsinghua.edu.cn}\\
\texttt{\{jamie.zhu,} \texttt{liuweiwen8,} \texttt{caiguohao1,} \texttt{zhangqi193,} \texttt{tangruiming,} \texttt{hexiuqiang1\}@huawei.com}
\vspace{1.5ex}
}

\thanks{$^\ast$ Both authors contributed equally to this work}
\thanks{$^\star$ Corresponding author}

\renewcommand{\authors}{Yi Li, Jieming Zhu, Weiwen Liu, Liangcai Su, Guohao Cai, Qi Zhang, Ruiming Tang, Xi Xiao, Xiuqiang He}

\begin{abstract}
The goal of recommender systems is to provide ordered item lists to users that best match their interests. As a critical task in the recommendation pipeline, re-ranking has received increasing attention in recent years.
In contrast to conventional ranking models that score each item individually, re-ranking aims to explicitly model the mutual influences among items to further refine the ordering of items given an initial ranking list. In this paper, we present a personalized re-ranking model (dubbed PEAR) based on contextualized transformer. PEAR makes several major improvements over the existing methods. Specifically, PEAR not only captures feature-level and item-level interactions, but also models item contexts from both the initial ranking list and the historical clicked item list. In addition to item-level ranking score prediction, we also augment the training of PEAR with a list-level classification task to assess users' satisfaction on the whole ranking list. Experimental results on both public and production datasets have shown the superior effectiveness of PEAR compared to the previous re-ranking models.

% (i) We model both  interactions to combine the advantages of both conventional ranking and re-ranking models. (ii) We not only leverage the contextual information from  as existing models have done, but also capture the inter-dependencies to  via sequence-to-sequence cross-attention. (iii) In addition to 

% 1) without leveraging historical items
% both feature-level and item-level interactions
% 2) entire list-level task the 
% 3) loss as multi-label learning problem

\end{abstract}








\maketitle

\input{sections/1_introduction}

\input{sections/2_related}
\input{sections/3_method}
\input{sections/4_experiment}
\input{sections/5_conclusion}
\section{Acknowledgments}
This work is partially supported by the National Natural Science Foundation of China (61972219), the Research and Development Program of  Shenzhen (JCYJ20190813174403598, SGDX20190918101201-696), the National Key Research and Development Program of China (2018YFB1800601), and the Overseas Research Cooperation Fund of Tsinghua Shenzhen International Graduate School (HW2021013).

\bibliographystyle{ACM-Reference-Format}
\balance
\bibliography{www}

\end{document}

%% file: sections/1_introduction.tex
\section{Introduction}
Recommender systems are playing an increasingly important role in our daily lives, serving a wide variety of  Web applications, such as e-commence, social networks, news feeds, etc. Due to the stringent latency requirements, industrial recommender systems typically follow a matching$\rightarrow$ranking$\rightarrow$re-ranking pipeline to gradually reduce the number of candidate items from millions to thousands, to hundreds, and finally to tens. As the final goal of recommender systems is to provide an ordered item list to users, re-ranking becomes a critical task and has received increasing attention in recent years~\cite{yin2016ranking,zhuang2018globally,ai2018learning,pei2019personalized}.

Most of the ranking models only consider the features of each user-item pair individually and learn a scoring function via a pointwise~\cite{cheng2016wide}, pairwise~\cite{cao2007learning} or listwise~\cite{xia2008listwise} loss. However, these models fail to explicitly model the mutual influences between items in the feature space. In contrast, re-ranking aims to model the scoring function by jointly encoding a list of items into contextualized representations, since the score of an item often depends on the other items placed in the same list~\cite{pei2019personalized}. This results in a more complex model architecture, which is usually used to refine the ordering of tens (e.g., top 30) of items in a given initial ranking list.

%Generating a list of items that meets user preferences is the main task of the recommendation model. Most of the research~\cite{xia2008listwise, taylor2008softrank,burges2005learning,burges2010ranknet,joachims2006training,cao2007learning, liu2009learning} focuses on how to obtain a high-quality recommendation list in the ranking stage. However, the ranking stage mainly focuses on the feature engineering and feature extraction of each item, which makes it unable to generate the optimal list. Therefore, the re-ranking model is proposed to optimize the final recommendation result from the perspective of the list. Owing to the ranking model, the items input to the re-ranking model are highly relevant to the user preference. So at the re-ranking stage, there is no need to consider filtering, but to focus on how to reorder these items to maximize the benefit of the list. Typically, mutual influences between items is a kind of important information.

% The re-rank model designed based on the output result of the ranking model can fully consider the important information neglected in the ranking stage, so as to achieve the optimal ranking result. 
% The main reasons are as follows: 1. The input data of the re-rank model is small enough 2. The input data of the re-rank model provides important information about the user's characteristics 3. The ordered list has the explicit of mutual influence between items.

Several pioneer research efforts~\cite{yin2016ranking,zhuang2018globally,ai2018learning,pei2019personalized} have been made to achieve this goal. Concretely, GlobalRank ~\cite{zhuang2018globally} and DLCM~\cite{ai2018learning} employ recurrent neural networks (RNNs) to encode the initial ranking list sequentially or bidirectionally. But RNNs have limited capacity in capturing pairwise item relationships. Later studies, such as PRM~\cite{pei2019personalized} and SetRank~\cite{pang2020setrank}, apply the self-attention mechanism in transformers to modeling inter-dependencies among items. Transformer architecture is more favorable due to its effectiveness in modeling interactions of any two items and its efficiency in parallelization. Although these models have focused on re-ranking modeling, they fail to take full advantage of the unique characteristic in recommender systems, that is personalization. Especially, many existing studies evaluate the re-ranking effectiveness using Web search datasets~\cite{pei2019personalized,pang2020setrank,ai2018learning,CRUM}. We argue that in recommender systems, the ranking of an item depends on not only the context of initial ranking list but also the context of historical clicked item list of a user. For example, if there are baby products and trendy clothing in the historical clicked list of a mother whose child is still infancy, then baby products may be ranked before the trendy clothing. Note that although user behavior sequence has been widely studied in other recommendation tasks, such as sequential recommendation for matching~\cite{Bert4Rec,SASRec} and click-through rate (CTR) prediction for ranking~\cite{zhou2019deep, chen2019behavior}, it is still unexplored for personalized re-ranking.

%For the ranking stage, the model only needs to perceive the user preference for these two categories of item so that the candidate items of the corresponding categories get higher scores than other categories. Applying such a strategy to the re-ranking phase obviously not meet the demand of re-ranking. 

To address the issue, in this work, we propose a personalized re-ranking model (dubbed PEAR) based on a novel contextualized transformer architecture, which comprises three main parts: feature-level interaction, item-level interaction, and multi-task training. More specifically, our PEAR model has made the following major improvements over the existing re-ranking models. 
%to fully leverage item contexts, including both the initial ranking list and the historical clicked item list. 

\begin{itemize}[leftmargin=*]
\item We model both feature-level and item-level interactions to combine the advantages of both ranking and re-ranking models. The feature interaction module can be substituted to any existing ranking network that allows to capture feature interactions, such as MLP~\cite{covington2016deep}, DCN~\cite{wang2017deep}, AFN~\cite{AFN}.
\item We not only leverage the contextual information within the initial ranking list as existing studies have done, but also capture the item inter-dependencies across the historical clicked item list via merged sequence-to-sequence cross-attention, which expands the context scope for improved personalization in re-ranking.

%but also mines the sequence information in historical behaviors that is highly consistent with the target of re-ranking.
\item We train PEAR through an item-level ranking score prediction task as well as a list-level classification task to assess users' satisfaction on the whole ranking list, which follows the multi-task training paradigm.
\end{itemize}

To evaluate the effectiveness of PEAR, we have conducted comprehensive experiments on a public benchmark dataset for micro-video recommendation~\cite{chen2018temporal} and a large-scale production dataset for news recommendation at Huawei. The results show the superiority of PEAR over the existing state-of-art methods.

%% file: sections/2_related.tex
\section{Related Work}

\textbf{Transformer-based ranking.} The great success of transformers in natural language processing (NLP) ~\cite{vaswani2017attention,kenton2019bert, liu2019roberta} draw much research attention from the recommendation domain. 
In recommendation tasks, user's historical behaviors with sequential structure is an important aspect for modeling.
Therefore, many studies~\cite{zhou2018deep,zhou2019deep,chen2019behavior,zhou2018atrank} use the self-attention mechanism in transformers to model relations among items. Zhou et al. ~\cite{zhou2018deep} introduces the target attention mechanism to adaptively assign importance weights to item embeddings in historical behaviors according to the target item for ranking. To better capture item relationships and sequential patterns underlying the historical behavior sequences, Chen et al. ~\cite{chen2019behavior} applies the transformer encoder to learn contextualized item representations of historical behaviors. Instead, our work leverages transformers for re-ranking.

\textbf{Re-ranking.} Different from the ranking stage where each item score is computed independently, re-ranking takes into account the mutual influences of items generated from a ranking model to refine their ordering. Several pioneer studies on re-ranking have been performed. ~\cite{zhuang2018globally} and ~\cite{ai2018learning} utilize RNNs to encode item-level sequential patterns into feature space. However, RNNs are limited in modeling pairwise relations directly, especially for long sequences. In contrast, PRM~\cite{pei2019personalized} introduces the self-attention mechanism to model inter-item dependencies for re-ranking. Also, it attaches a personalized vector from the ranking model to each item in order to enhance the personalization effect in re-ranking. In addition,~\cite{liu2020personalized} proposes a graph neural network based model to capture item relationships. Qiu et al.~\cite{Rerank_Augment} investigate the use of data augmentation to mitigate the imbalance issue in re-ranking. Different from these models, our work leverages contextualized transformers to incorporate historical behaviors for re-ranking. 

% And we design a effective mechanism to encode both mutual influence among items in initial list and item inter-dependencies across the historical clicked item list.

%% file: sections/3_method.tex
% \begin{figure*}[!pt]
% 	\centering
% 	\includegraphics[width=0.8\linewidth]{figures/Model.pdf}%\vspace{-1ex}
% 	\caption{Overview of PEAR.}
% 	\label{fig:model}
% \end{figure*}

\begin{figure}[!pt]
	\centering
	\includegraphics[width=\linewidth]{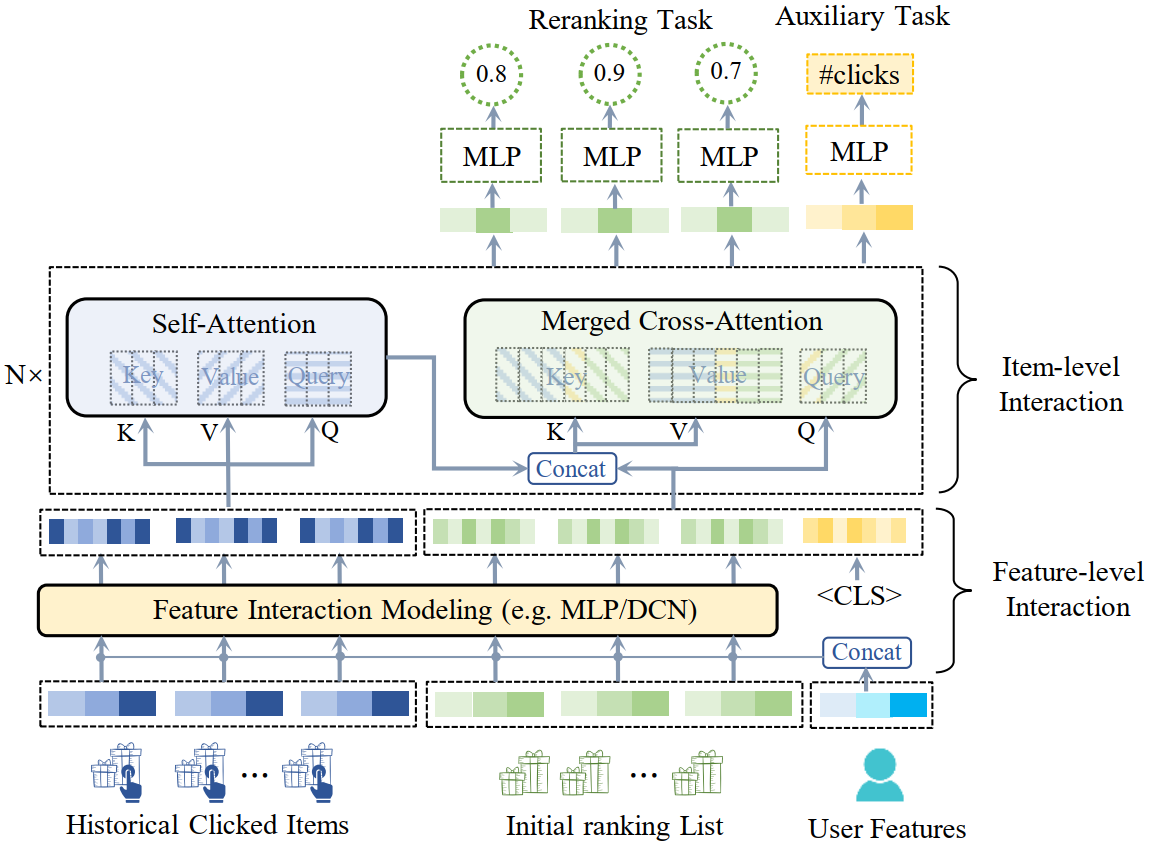}
	\caption{The model architecture of PEAR.}
	\label{fig:model}
	\vspace{-1ex}
\end{figure}

\section{Methodology}
In this section, we describe the PEAR model architecture as illustrated in Figure \ref{fig:model}. There are three main parts in the framework: feature-level interaction, item-level interaction and multi-task training. In what follows, we describe each part in details.

%First, feature-level interaction capture complex interaction across features of each item representation. Second, the item-level interaction part considers the dependencies between items in the whole list. To achieve this goal, we propose a contextualized Transformer. Third, in order to better train our model, we design two training tasks. In the following sections, we penetrate the details of three parts.

\subsection{Feature-level Interaction}
Learning interactions of raw features is important as indicated in existing studies for ranking~\cite{wang2017deep,guo2017deepfm}. To fuse features (e.g., gender, city, category) of each user-item pair, we propose a feature interaction module to capture complex feature interactions and thus obtain contextualized item representations. In recent years, many effective feature interaction networks have been proposed, such as  DeepFM~\cite{guo2017deepfm}, DCN~\cite{wang2017deep}, and AFN~\cite{AFN}, to capture useful feature interactions. While it is generally applicable to employ any of these networks as our feature interaction module, in this work, we use a two-layer MLP for simplicity. 

%Moreover, we can directly use the trained ranking model to fine-tune according to the re-ranking target. 

Typically, both user features and item features comprise multi-field categorical values (e.g., the city field contains Beijing/Shanghai/$\cdots$). It is common to apply embedding layers to map these highly sparse categorical features into dense feature embeddings (as detailed in~\cite{FuxiCTR}). After that, we could obtain a flattened feature vector for each user ($f_u \in \mathbb{R}^{d_u}$) and item ($f_i \in \mathbb{R}^{d_i}$), where $d_u$ and $d_i$ denote flattened dimensions. To get fused user-item representations, we first concatenate the user feature vector and every item feature vector in the historical item list $B=[b_1, b_2, ..., b_n]$ and the initial ranking list $S=[s_1,s_2,...,s_m]$, which produces the representation matrix $X \in \mathbb{R}^{(d_u+d_i) \times(n+m)}$. Then, we feed X into a two-layer MLP as follows:
\begin{equation}
\begin{aligned}
Z &= W_2\,\sigma(W_1X+b_1)+b_2~,
\end{aligned}
\end{equation}
where $W_1, W_2, b_1, b_2$ are learned kernel and bias weights. $\sigma$ denotes the activation function (e.g., ReLU). $Z\in \mathbb{R}^{d\times (n+m)}$ outputs the fused representations in $d$-dimension. It is worth noting that our feature interaction module (i.e., 2-layer MLP) works similarly to the feedforward network (FFN) in transformers~\cite{vaswani2017attention}. In addition, to facilitate the subsequent list-level task, we add a special classification token (i.e., CLS) with learnable parameters in $\mathbb{R}^d$ at the end of the initial ranking list. It is inspired by Bert~\cite{kenton2019bert}, and is useful in aggregating the list-level representation of the entire initial ranking list. Finally, we obtain $Z_B \in \mathbb{R}^{d\times n}$ and $Z_S\in \mathbb{R}^{d\times (m+1)}$ as the output representations of the historical item list and the initial ranking list (with CLS token), respectively.

%Taking inspiration from the BERT, we place a special classification token (<CLS>) at the end of the initial list. The final hidden state of this token aggregates , which can be used for the subsequent list-level classification task. 
%Thus, the completed output matrix of this module can be denoted as $O \in \mathbb{R}^{(n+m+1)\times d}$.
% Thus, the completed output matrix of this module can be formulate as $E_f = [x^{(2)}_{h_1},...,x^{(2)}_{h_n}, x^{(2)}_{s_1},...,x^{(2)}_{s_m}, r_{cls}]$. $r_{cls}$ is a randomly generated d-dimensional vector.

\subsection{Item-level Interaction}
Several pioneer research efforts \cite{bello2018seq2slate, zhuang2018globally,ai2018learning,pei2019personalized} have been made to model the mutual influences between item pairs. Although they introduce different network structures (e.g., RNNs, self-attention) for item relation modeling, all of them only consider the context of the initial ranking list, largely neglecting the potential influences by the historical clicked item list. However, historical user behaviors contain rich and fine-grained user interests that are equally important for contextualized item modeling in re-ranking.

In this work, we employ the popular encoder-decoder structure~\cite{vaswani2017attention} that has been widely applied in NLP to capture the item-level interactions not only within the initial ranking list and within the historical item list, but also across the two lists. As shown in Figure \ref{fig:model}, the item-level interaction module consists of a stack of N blocks, each comprising a self-attention layer and a merged cross-attention layer. For efficiency, we set $N=1$ in our experiments. 

First, a self-attention layer is employed to model fine-grained user interests in the historical behaviors, which could provide more informative contexts for learning item interactions across lists. Specifically, we use the self-attention in~\cite{vaswani2017attention} as follows:
\begin{equation}
\begin{aligned}
H_{B} &= Softmax\bigg(\frac{(W_QZ_B)^T(W_KZ_B)}{\sqrt{d_h}} \bigg)(W_VZ_B)^T~,
\end{aligned}
\end{equation}\label{eq:sa}
where $W_Q, W_K, W_V \in \mathbb{R}^{d_h \times d}$ are learnable query, key, and value parameters. $H_B \in  \mathbb{R}^{n \times d_h}$ represents the output item representations in historical item list. 

Then, a merged cross-attention layer is applied to model the item interactions within the initial ranking list and across the two lists simultaneously. For this purpose, a straightforward way is to apply a self-attention sub-layer and a cross-attention sub-layer accordingly, as the transformer decoder does in~\cite{vaswani2017attention}. But we further merge the two sub-layers together for  better computation efficiency in GPUs. More specifically, given $H_B$ and $Z_S$ as inputs, the merged cross-attention is formulated as follows:
\begin{equation}
\begin{aligned}
H_S = Softmax\bigg(\frac{(W_qZ_S)^T[W_{k1}H_B^T, W_{k2}Z_S]}{\sqrt{d_h}} \bigg)[W_{v1}H_B^T, W_{v2}Z_S]^T
\end{aligned}
\end{equation}\label{eq:att}
where $W_q, W_{k2}, W_{v2}\in \mathbb{R}^{d_h \times d}$ and $W_{k1}, W_{v1} \in \mathbb{R}^{d_h \times d_h}$ are all learnable parameters. [$\cdot$] denotes the concatenation operation. In this way, we use a single merged cross-attention layer to simultaneously capture item interactions within $Z_S$ and across $Z_S$ and $H_B$. 

Note that our PEAR model comprises a feature-level interaction module and an item-level interaction module, corresponding to FFN and multi-head self-attention layers respectively in a vanilla transformer~\cite{vaswani2017attention}. Therefore, we view PEAR as a contextualized transformer.

\subsection{Multi-task Training}
The output $H_S$ goes through a one-layer MLP followed by sigmoid activations for click prediction. It is worth noting that softmax loss does not work well since multiple positives may exist in the initial ranking list, making the sum of ground-truth probabilities larger than 1. Instead, we model the task as a multi-label learning problem using binary cross-entropy loss, which is defined as:  
\begin{equation}
\mathcal{L}_{m} =  \sum{\Big(y_tlog(\hat{y}_t)+(1-y_t)log(1-\hat{y}_t)\Big)}
\end{equation}
where $y_t \in \{0,1\}$ denotes whether the user has clicked the target item. $\hat{y}_t$ is the predicted click probability of the target item. 

Considering that the above loss only considers each single item in isolation, we introduce an auxiliary task, which is a list-level task to predict the number of clicks in each initial list. For simplicity, in this work, we also formulate it as a binary classification problem with the following loss:
\begin{equation}
\mathcal{L}_{aux} = \sum{\Big(y_{aux}log(\hat{y}_{aux})+(1-y_{aux})log(1-\hat{y}_{aux})\Big)}
\end{equation}
where $y_{aux}$ represents whether the initial ranking list contains positive items. $\hat{y}_{aux}$ is the predicted probability from the CLS token. At last, we combine the two loss functions as multi-task learning:
\begin{equation}
\mathcal{L} = \mathcal{L}_{m} + \alpha \mathcal{L}_{aux}
\end{equation}
where $\alpha$ is a hyper-parameter to adjust the effect of $\mathcal{L}_{aux}$. Typically, we set $\alpha=1$.

%% file: sections/4_experiment.tex
\section{Experiments}
% In this section, we conduct experiments on two datasets to demonstrate the effectiveness of our proposed PEAR. %Then we further design the ablation study to investigate the effectiveness of several components.
\subsection{Experimental Settings}

\begin{table*}[!tph]\large
\vspace{-4ex}
\caption{Performance comparison of PEAR to other re-ranking methods.}
\centering
\small
\resizebox{0.95\textwidth}{!}{
\begin{tabular}{c|cccc|cccc}
\hline
\multirow{2}{*}{Model} & \multicolumn{4}{c|}{MicroVideo-1.7M} & \multicolumn{4}{c}{News-Dataset} \\ \cline{2-9}
 & gAUC@20 & gAUC@30 &nDCG@20 &nDCG@30 & gAUC@20 & gAUC@30 &nDCG@20 & nDCG@30 \\ \hline
 DCN (base) & 0.5342 & 0.5405 & 0.5349 & 0.6585 & 0.6014 & 0.6097 & 0.2557 & 0.2601 \\ \hline
 DLCM & 0.5604 & 0.5665 & 0.5571 & 0.6711 & 0.6040 & 0.6115 & 0.2564 & 0.2608 \\
 SetRank & 0.5483 & 0.5537 & 0.5445&  0.6620& 0.6017&  0.6099&  0.2555& 0.2602 \\
 PRM & \underline{0.5629} & \underline{0.5680} & \underline{0.5579} & \underline{0.6714} & \underline{0.6059} & \underline{0.6125} & \underline{0.2572} & \underline{0.2616} \\
PEAR & \textbf{0.5681} &  \textbf{0.5741} &  \textbf{0.5640} &  \textbf{0.6755} &  \textbf{0.6172} &  \textbf{0.6230} &  \textbf{0.2599} &  \textbf{0.2641}\\ \hline
%  & Improv.  & +27.0\%  & +23.2\% & +20.7\% & +16.4\% & +12.6\%  & +7.6\% & +13.5\% & +7.1\% \\ \hline
\end{tabular}
}
\label{Experimental_results}
\vspace{-1ex}
\end{table*}

\subsubsection{Dataset Description}
Existing works conduct experiments on Web search datasets~\cite{pei2019personalized,pang2020setrank,ai2018learning,CRUM}. However, these datasets do not contain user historical behaviors required by our work. Thus, we employ another two real-world datasets in our experiments:  
\begin{itemize}[leftmargin=*]
\item \textbf{MicroVideo-1.7M}: This is a million-scale dataset collected from a popular micro-video recommender system in China, which has 12,737,619 interactions that 10,986 users have made on 1,704,880 micro-videos~\cite{chen2018temporal}. We follow the existing train-test data splits.
\item \textbf{News-Dataset}: The private dataset contains click logs of three hours, randomly sampled from Huawei's production news recommender systems. It is composed of 17,203,522 interactions between 1.8M users and 37K news.
\end{itemize}

\subsubsection{Evaluation Metrics} We use the following common metrics for evaluating re-ranking performance~\cite{ai2018learning, zhou2018deep}, including nDCG@K and gAUC@K, where K=$20, 30$ following ~\cite{pei2019personalized}.

%in our evaluations to compare different methods. \textbf{i}) normalized Discounted Cumulative Gain (nDCG). \textbf{ii}) gAUC. 
% gAUC is a variation of user weighted AUC that measures the goodness of intra-user. It is defined as follows:
% \begin{equation}     
%     \label{eq:GAUC}
%     gAUC = 
%     \frac{\sum_{i=1}^n \# impression_{i} \times AUC_{i}}{\sum_{i}^n \# impression_{i}}
% \end{equation}
% where $n$ and $\#impression_{i}$ represent the number of users and impressions, respectively. $AUC_{i}$ is the AUC corresponding to the $i$-th user.
%More specifically, we report results of nDCG@K and gAUC@K, where K=20 and K=30 are used. For a fair comparison, the maximum length of initial list in our experiment is 30, the same as 

% More specifically, we report results of gAUC@20, gAUC@30, nDCG@20 and nDCG@30. For a fair comparison, the maximum length of initial list in our experiment is 30, the same as ~\cite{pei2019personalized}.

\subsubsection{Initial ranker and baselines}
To generate an initial ranking list for the re-ranking model, we select DCN, a learning to rank method that achieves great success in industry. And we compare our proposed PEAR against several baseline re-ranking models including DLCM~\cite{ai2018learning}, PRM~\cite{pei2019personalized} and SetRank~\cite{pang2020setrank}.

% \begin{itemize}[leftmargin=*]

% \item \textbf{DLCM}~\cite{ai2018learning}: DLCM applies GRU to encode local ranking context, and rerank items with a local ranking function.
% \item \textbf{PRM}~\cite{pei2019personalized}: PRM uses a transform structure to encode the information of all items in the initial list, and learn user preference by a personalized module.
% \item \textbf{SetRank}~\cite{pang2020setrank}: SetRank employs multi-head self- attention mechanism to achieve cross-document interactions and permutation-invariant.
% \end{itemize}

%%%%%%%%%%%%%%%%%%%%%%%%%%%%%%%%

%%%%%%%% Huawei-Dataset %%%%%%%%%

\subsubsection{Implementation details} We use Tensorflow for model implementation and apply the Adam~\cite{kingma2014adam} optimizer. The default length of the historical behavior is 128. The MLP used for feature-level interaction is set to [500, 500]. We set the dropout rate to 0.1, batch size to 200 for MicroVideo-1.7M and 400 for News-Dataset. We tune the learning rate from \{1e-3, 1e-4, 5e-5, 1e-5\}. The number of heads is set 1 for MicroVideo-1.7M and 2 for News-Dataset. The hidden dimensionality $d_h$ is 500. We  apply early stopping (patience$=2$) to avoid overfitting during training. We also tune the baseline models exhaustively to get reasonable results. 

% \begin{table*}[!tph]\large
% \caption{Comparison of competing methods and PEAR on different datasets.}
% \centering
% \resizebox{0.95\textwidth}{!}{
% \begin{tabular}{c|cccc|cccc}
% \hline
% \hline
% \multirow{2}{*}{Model} & \multicolumn{4}{c|}{MicroVideo-1.7M} & \multicolumn{4}{c}{Huawei-Dataset} \\ \cline{2-9}
%  & gAUC@20 & gAUC@30 &nDCG@20 &nDCG30 & gAUC@20 & gAUC@30 &nDCG@20 & nDCG@30 \\ \hline
%  DCN (ranking) & 0.5342 & 0.5405 & 0.5349 & 0.6585 & 0.6014 & 0.6097 & 0.2557 & 0.2601 \\ \hline
%  DLCM & 0.5604 & 0.5665 & 0.5571 & 0.6711 & 0.6040 & 0.6115 & 0.2564 & 0.2608 \\
%  SetRank & 0.5483 & 0.5537 & 0.5445&  0.6620& 0.6017&  0.6099&  0.2555& 0.2602 \\
%  PRM & \underline{0.5629} & \underline{0.5680} & \underline{0.5579} & \underline{0.6714} & \underline{0.6059} & \underline{0.6125} & \underline{0.2572} & \underline{0.2616} \\
% PEAR & \textbf{0.5681} &  \textbf{0.5741} &  \textbf{0.5640} &  \textbf{0.6755} &  \textbf{0.6172} &  \textbf{0.6230} &  \textbf{0.2599} &  \textbf{0.2641}\\ \hline
% \hline
% %  & Improv.  & +27.0\%  & +23.2\% & +20.7\% & +16.4\% & +12.6\%  & +7.6\% & +13.5\% & +7.1\% \\ \hline
% \end{tabular}
% }
% \label{Experimental_results}
% \end{table*}

\subsection{Performance Comparison}
The overall performance of our baselines and PEAR are shown in Table \ref{Experimental_results}. For each dataset, we report the result of one ranking model DCN, three re-ranking sota baselines and PEAR. The best performance are highlighted in bold. 

Results show that PEAR consistently outperforms all baselines across the public dataset and the industry dataset in all cases. Specifically, it achieves improvements over the strongest baselines in terms of gAUC@30 by 1.07\%, 1.71\%, in MicroVideo-1.7M and Huawei-Dataset, respectively. As for nDCG, the highest nDCG@20 of baselines reported on MicroVideo-1.7M is 0.5579. While our PRM can reach 0.5640, which achieve 1.09\% improvements.
This validates the effectiveness of our architecture for modeling historical behaviors in the re-ranking scenario. We owe this to the modeling of the dependency between the initial list and the historical list. This kind of explicit modeling between sequences can dig out the information in the historical behaviors that directly guides the reordering of the initial list.
It is worth noting that PRM that does not regard the historical behaviors as a sequence for modeling is the most competitive baseline. One possible reason is that PRM benefits a lot from its personalized vector generated from historical behavior through a pre-training model, and this information is not used in other baselines. Nonetheless, because of the way it handles historical behaviors, the improvements of PRM brought by it is limited. In addition, noted that SetRank performs worst in the baseline models, which may be due to the attention rank loss adopted by SetRank is not suitable for recommendation task.

% The most competitive baseline is PRM, which generates a personalized vector from historical behavior through a pre-training model as part of the input. Obviously, other re-ranking baselines that do not incorporate information from historical behavior are inferior to PRM. Nonetheless, PRM not modeling historical behaviors as sequence.

\begin{table}[!t]\large
%\vspace{-1ex}
\caption{Ablation on multi-task training, MicroVideo-1.7M.}

\centering
\resizebox{1.0\columnwidth}{!}{

\begin{tabular}{ccccc}
\hline
Model & gAUC@20 & gAUC@30  & nDCG@20  &nDCG@30   \\ \hline

w/o auxiliary task & 0.5606 & 0.5672 & 0.5626 & 0.6747 \\
w/ auxiliary task & \textbf{0.5681} & \textbf{0.5741} & \textbf{0.5640} & \textbf{0.6755} \\
\hline
\end{tabular}
}
\label{table:aux}
\vspace{-3ex}
\end{table}

\begin{figure}[!t]
	\centering
	\vspace{-2ex}
	\includegraphics[width=\linewidth]{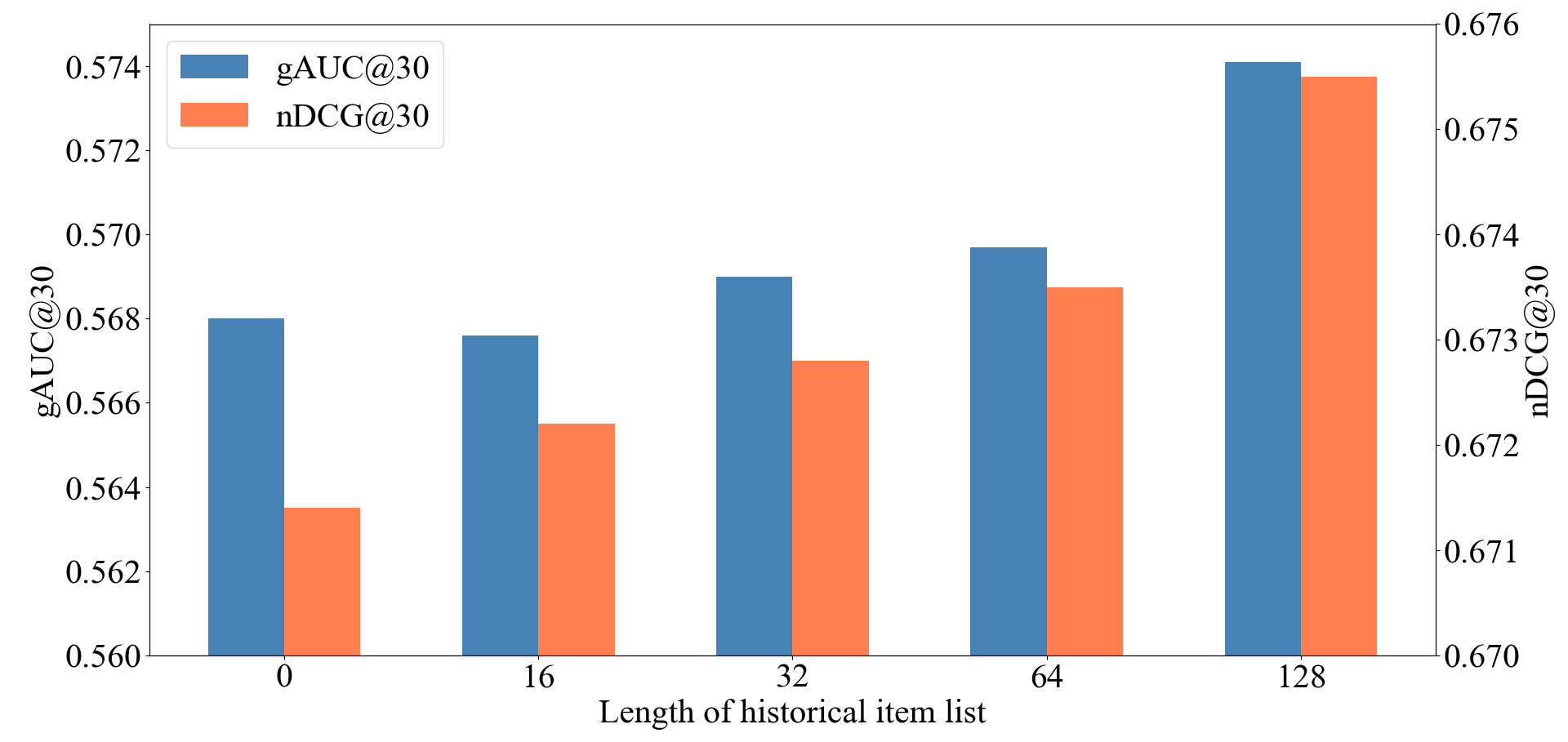}%\vspace{-1ex}
	\caption{Results of the PEAR with different history length on MicroVideo-1.7M datasets.}
	\vspace{-2ex}
	\label{fig:ablation}
	
\end{figure}

\subsection{Ablation Study}
% We perform ablation studies on PEAR by showing how auxiliary task and the length of user click behavior affect its performance.
\subsubsection{Effect of auxiliary task}
To prove the efficiency of the auxiliary task, we compare the performance of our model without and with auxiliary task on MicroVideo-1.7M dataset. Results are shown in Table \ref{table:aux}. According to the results, the nDCG@30 drops from 0.5741 to 0.5672 when removing the auxiliary task. This verifies the effect of the global supervision signal of the list.

\subsubsection{Effect of length of historical item list}
Figure \ref{fig:ablation} shows the performance comparison of PEAR with different history lengths on the MicroVideo-1.7M dataset. Due to the space limitation, the results of gAUC@30 and nDCG@30 are selected for reporting. Obviously, as the length of historical sequence increases, the performance of our model is gradually improving. When the length of the historical sequence changes from 64 to 128, the performance improvement is most significant. This justifies the effectiveness of the sequence signal of historical behaviors in the re-ranking stage.

Therefore, the above ablations show that each component in PEAR is effective in refining the order of the initial ranking list.
%\begin{itemize}[leftmargin=*]
%\item When PEAR only uses a relatively short history sequence such as 16, it also has comparable performance to PRM. As long as the historical sequence is slightly extended to 32, PEAR shows its advantages. 
%This is because the length of the initial list is 30, and the history sequence shorter than this length can only play a limited role in refining the whole list. 
%This indicates that compared to PRM extracting personalized vectors from the historical sequence, the explicit modeling of dependency between historical clicked items and items in the candidate list is beneficial to optimize the entire list.
%\item Focusing on the variants of PEAR, shown in the last four positions in the figure. As the length of historical sequence increases, the performance of the PEAR model is gradually improving. When the length of the historical sequence changes from 64 to 128, the performance improvement is most significant. Based on this, we adopt the history length of 128 in our experiments. This justifies the effectiveness of the sequence signal of historical behavior in the re-ranking stage. 
%\end{itemize}

%% file: sections/5_conclusion.tex
\section{Conclusion}
In this paper, we focus on personalized re-ranking with a contextualized transformer model, namely PEAR. In contrast to existing re-ranking models, our work makes three main improvements, including (i) modeling both feature-level and item-level interactions, (ii) capturing the contextual information within initial ranking list as well as across historical clicked item list, and (iii) applying multi-task learning to augment the training of PEAR with a list-level classification task. Extensive experiments have been conducted on both a public micro-video recommendation dataset and a production news recommendation dataset from Huawei to validate the effectiveness of our PEAR model. Future work might include improving model efficiency further and considering long-tail users (whose historical item list is too short) better. 

%the role of historical behaviors in the re-ranking task. A few studies incorporate a compressed representation vector of historical behaviors into the raw features of items in the initial list. Unfortunately, the special modeling of the inter-sequence context is ignored. We design a novel model named PEAR that can model the inter-sequence effect of the initial list and the historical behaviors while capturing the mutual influence of the initial list, using an effective architecture based on attention mechanism. Besides, a loss that can supervise the global prediction of the initial list brings further performance improvements to the model. 